\documentclass[twocolumn,prl,aps,longbibliography,superscriptaddress,floatfix,tightenlines]{revtex4-2}
\usepackage{amsmath}
\usepackage{amssymb}
\usepackage{mathrsfs}
\usepackage{amsfonts}
\usepackage{epsfig}
\usepackage{bm}
\usepackage{times}
\usepackage{lineno}
\usepackage[T1]{fontenc}
\usepackage[dvips]{color}
\usepackage{braket}
\usepackage[colorlinks,bookmarks=false,citecolor=magenta,linkcolor=magenta,urlcolor=blue]{hyperref}
\definecolor{darkblue}{rgb}{0.0, 0.0, 0.6}

\begin{document}

\title{%Sensitivity and efficiency of
Non-Hermitian %sensors
sensing from the perspective of post-selected measurements}
\author{Neng Zeng}
\affiliation{School of Physics and Optoelectronics, South China University of Technology, Guangzhou 510640, China}

\author{Tao Liu}
%\email{yuranzhang@scut.edu.cn}
\affiliation{School of Physics and Optoelectronics, South China University of Technology, Guangzhou 510640, China}

\author{Keyu Xia}
\affiliation{College of Engineering and Applied Sciences, National Laboratory of Solid State Microstructures, Nanjing University, Nanjing 210023, China}
\affiliation{Shishan Laboratory, Suzhou Campus of Nanjing University, Suzhou 215000, China}

\author{Yu-Ran Zhang}
\email{yuranzhang@scut.edu.cn}
\affiliation{School of Physics and Optoelectronics, South China University of Technology, Guangzhou 510640, China}
%\affiliation{Theoretical Quantum Physics Laboratory, RIKEN Cluster for Pioneering Research, Wako-shi, Saitama 351-0198, Japan}

\author{Franco Nori}
%\affiliation{Theoretical Quantum Physics Laboratory, RIKEN Cluster for Pioneering Research, Wako-shi, Saitama 351-0198, Japan}
\affiliation{Center for Quantum Computing, RIKEN, Wakoshi, Saitama 351-0198, Japan}
\affiliation{Physics Department, University of Michigan, Ann Arbor, Michigan 48109-1040, USA}
\date{\today}
\begin{abstract}
By employing the Naimark dilation, we establish a fundamental connection
between non-Hermitian quantum sensing and post-selected measurements.
The sensitivity of non-Hermitian quantum sensors is determined by the effective quantum Fisher information (QFI), which incorporates the success probability of post-selection.
We demonstrate that non-Hermitian sensors cannot outperform their Hermitian counterpart when all information is harnessed, since the total QFI for the extended system constrains the effective QFI of the non-Hermitian subsystem.
Moreover, we quantify the efficiency of non-Hermitian sensors with the
 ratio of the effective QFI to the total QFI, which can be optimized
 within the framework of post-selected measurements with minimal experimental trials.
Our work provides a distinctive theoretical framework for
investigating non-Hermitian quantum sensing and designing noise-resilient
quantum metrological protocols.
\end{abstract}

\maketitle
 \emph{Introduction.---}%
Quantum metrology leverages quantum coherence and entanglement to enhance sensitivity and accuracy in measuring physical quantities \cite{giovannetti2006quantum,giovannetti2011advances,degen2017quantum}, with promising applications in
various fields of modern science and technology \cite{kitching2011atomic,aigner2008long,hosten2008observation}. The achievements of modern quantum physics have also introduced novel frameworks and protocols for quantum-enhanced metrology \cite{Pezze2017,Zhang2018,Brenes2020,Zhao2020,Desaules2022,Li2023b}.
Recently, several metrological schemes based on non-Hermitian physics have been theoretically
proposed and experimentally demonstrated, e.g., enhanced sensing near  exceptional points (EPs)
\cite{hokmabadi2019non,yu2020experimental,lai2019observation,hodaei2017enhanced,chen2017exceptional,wiersig2014enhancing,wang2022boosting,liu2016metrology,rosa2021,tang2023,li2023,xu2024single,Ruan2025}.
%For instance, EP-based sensors can realize enhanced sensing with arbitrary precision since the resonant frequencies become strongly dependent on perturbation near the  exceptional point (EP), and they have been demonstrated in several systems \cite{hokmabadi2019non,yu2020experimental,lai2019observation,hodaei2017enhanced,chen2017exceptional,wiersig2014enhancing,liu2016metrology,wang2022boosting,liu2016,rosa2021,tang2023,li2023}.
Due to the divergence of the susceptibility in the vicinity of EPs,
 EP-based sensors have been theoretically predicted to realize enhanced sensing %with arbitrary precision
\cite{liu2016metrology}.
The  EP-based sensors can be realized in $\mathcal{PT}$-symmetric systems with two EPs at the phase transition points of $\mathcal{PT}$-symmetry breaking, which have been demonstrated in open systems with loss and gain \cite{ruter2010observation,peng2014parity,chang2014parity,ozdemir2019parity}.
%with  arbitrary precision
Several experiments have implemented EP-based sensing with %the viewpoint and curiously fine the enhance of transduction and noise are exactly cancel each other out near EP \cite{wang2020petermann}, but others have demonstrated
an enhanced signal-to-noise ratio (SNR) \cite{zhang2019quantum,wang2020petermann,kononchuk2022exceptional,peters2022exceptional}.
However, some works doubt whether EP-based sensors can improve the fundamental sensitivity limits in presence of noise  \cite{langbein2018no,wiersig2020prospects,loughlin2024exceptional,Zheng2025,Naikoo2023}.
In addition, another non-Hermitian sensing scheme independent of EPs has been proposed, which is predicted to be robust against some specific forms of noise \cite{chu2020quantum,xiao2024non}.
Whether non-Hermitian sensors can offer an advantage in estimation  sensitivity over conventional Hermitian sensors is still a controversial topic \cite{wiersig2020review}, despite
studies on the metrological limits of
non-Hermitian sensing \cite{lau2018fundamental,Bao2021,ding2023fundamental}.

%Although some theoretical works give fundamental limits of non-Hermitian quantum sensors \cite{lau2018fundamental,ding2023fundamental},  is superior to normal sensors is still a topic of debate \cite{wiersig2020review}.
Non-Hermitian sensing involves projection-value measurements (PVMs) merely on the quantum subsystem that exchanges energy with its environment. The quantum open system dynamics are more rigorously described with approaches, such as the Kraus representation and Lindblad formalism  \cite{bender2007making,ashida2020non,bender2024}. For noisy quantum parameter estimation (QPE), the minimum achievable statistical uncertainty  is determined by the quantum Fisher information (QFI) and the Cram\'{e}r-Rao bound (CRB)
\cite{braunstein1994statistical,Ma2011}. It relates to
the minimum QFI corresponding to a unitary evolution of the enlarged system \cite{escher2012quantum,escher2011general}.
Thus, from the perspective of quantum information science \cite{toth2014quantum}, the metrological resources from both
the open system and its environment should be considered when comparing the performances of
non-Hermitian sensors and their Hermitian counterparts \cite{ding2023fundamental}.
This problem is in analog with the measurement sensitivity of weak-value-amplification (WVA) technique \cite{aharonov1988result,Kofman2012,dressel2014colloquium,Pang2014,Dressel2010}, which greatly
improves the SNR by discarding  most detection trials.  WVA can be described as
POVMs on the sensor subsystem, interacting with an ancillary system.
With the Naimark extension theorem and its inference \cite{dalla2019naimark,beneduci2020notes}, a non-Hermitian system can be dilated to
a larger Hermitian system followed by post-selecting the ancilla state \cite{huang2019simulating,gunther2008naimark,wu2019observation,Minganti2020}.
Therefore, non-Hermitian sensing can be understood in the framework of the QPE with post-selected measurements, of which the measurement sensitivity in the presence of technical noise have been widely discussed
\cite{ferrie2014weak,knee2014amplification,zhang2015precision,harris2017weak,brunner2010,starling2009,pang2015,jordan2014,knee2016,Tanaka2013,kim2022,Jozsa2007,wang2016experimental,Vaidman2017,Aharonov1996,Lyons2015}.

Here, we investigate the sensitivity of non-Hermitian sensors by
establishing a fundamental connection between non-Hermitian sensing and
QPE with post-selected measurements.
%relating it to the QPE with post-selected measurements.
%methods of quantum metrology.
Analogous to  WVA, we show the enhancement of QFI corresponding to non-Hermitian sensors and prove that the effective QFI,
when considering the success probability, is not larger than the total QFI
of the Naimark-dilated Hermitian system.
%the corresponding between non-Hermitian sensors and post-selection scheme.
Therefore, when amounting for the neglected resources from the environment,
 non-Hermitian sensor cannot outperform their Hermitian counterparts,  regardless of the specific form of decoherence or the choice of probe states.
Specifically, we analyze three specific types of non-Hermitian sensing proposals, including a pseudo-Hermitian sensor and two EP-based sensors.
%which were predicted to perform non-Hermiticity-enhanced sensing.
Moreover, we show that the efficiency
of non-Hermitian sensing, evaluated by the ratio of the effective QFI to the total QFI, can be optimized through post-selection protocols requiring minimal experimental trials.
Our work provides a comprehensive understanding of non-Hermitian quantum sensing and is useful for exploring practical and efficient quantum metrology schemes against technical noise.

%On the basis of previous results of post-selection and WVA, we study the QFI of the non-Hermitian sensors and their Hermitian counterparts, where the results imply non-Hermitian sensors is suboptimal and cannot outperform their Hermitian counterparts.
%Specifically, the effective information of the non-Hermitian sensors will be no larger than the all information of the environment resource which maintains the target system.
%Subsequently, we have scrutinized various types of non-Hermitian sensing proposals. We have %discussed two EP-based sensors, one of which is $\mathcal{PT}$-symmetric and the other is not.
%	Moreover, a single-qubit pseudo-Hermitian system without EP has been considered, and all of these three sensing proposals are in agreement with our predictions.

\emph{Quantum parameter estimation (QPE).---}%
QPE aims to estimate an  unknown parameter $\theta$, imprinted on a quantum state $\rho_\theta$. Measurements in terms of POVMs $\{\hat{E}(x)\}$ are performed on $\rho_\theta$, yielding outcomes $\{x\}$ with probabilities $P(x|\theta)=\textrm{tr}[\rho_\theta\hat{E}(x)]$.
For an unbiased estimator $\hat{\theta}_{\textrm{est}}$, the sensitivity of QPE is evaluated with its variance: $(\delta \theta)^2\equiv{\langle \hat{\theta}_{\textrm{est}}^2 \rangle-\langle \hat{\theta}_{\textrm{est}}\rangle^2}$, which is lower bounded by the CRB: $(\delta \theta)^2\ge1/{\nu F(\theta)}$, with $\nu$ being the repetition number of measurements ~\cite{helstrom1969,wiseman2009quantum}.
Here, $F(\theta)\equiv\int \! dx \;  [\partial_\theta P(x|\theta)]^2/P(x|\theta)$ denotes the Fisher information (FI), and the CRB can be approximately saturated for $\nu \to \infty$.
%For the appointed state $\rho_\theta$,
The QFI denotes the maximum FI over all possible POVMs, i.e.,
$F_Q(\theta) \equiv \max_{\{\hat{E}(x)\}} F(\theta)$, which
%In addition, the QFI
can be expressed with the symmetric logarithmic derivative %(SLD)
as $F_Q(\theta)=\textrm{tr}(\rho_\theta\hat{\mathcal{L}}_\theta^2)$, with $\hat{\mathcal{L}}$ satisfying $\partial_\theta\rho_\theta=(\rho_\theta\hat{\mathcal{L}}_\theta+\hat{\mathcal{L}}_\theta\rho_\theta)/2$~\cite{braunstein1994statistical}.
For a pure state $\rho_\theta=\ket{\psi_\theta}\bra{\psi_\theta}$, the QFI can be simplified as $F_Q(\theta)=4(\langle  \partial_\theta\psi_\theta|\partial_\theta\psi_\theta\rangle+|\langle\psi_\theta|\partial_\theta\psi_\theta\rangle|^2)$~\cite{giovannetti2011advances,giovannetti2006quantum}.

\begin{figure}[t]
	\centering
	\includegraphics[width=0.48\textwidth]{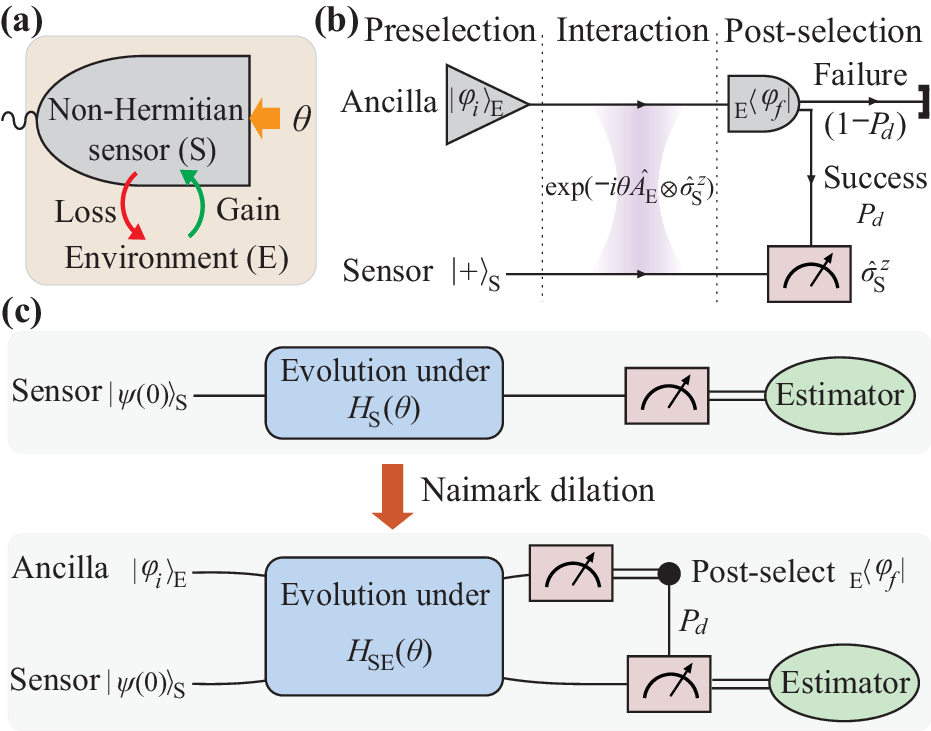}\\
	\caption
	{(a) Non-Hermitian sensor $\textrm{S}$ detects an unknown parameter $\theta$, by coupling to an environment $\textrm{E}$ with gain and/or loss.
	(b) Schematic of WVA, as a representative post-selected measurement strategy for estimating $\theta$. The pre- and post-selected states are $|\psi_i\rangle_\textrm{E}$ and $|\psi_f\rangle_\textrm{E}$, respectively, with success and rejection probabilities $P_d$ and $(1-P_d)$, respectively.
	(c) Quantum circuits for the non-Hermitian sensing protocol
	and its Hermitian counterpart. The non-Hermitian Hamiltonian $H_\textrm{S}(\theta)$ and the extended Hamiltonian $H_{\textrm{SE}}(\theta)$  are interconnected through the Naimark dilation theorem.}
	%system can be regarded as the post-selection process on their Hermitian counterparts. In the sketch map of post-selection, $\ket{\Psi}$ represents the state of Hermitian counterpart, witch exists in extended Hilbert space $\mathcal{H}_s\otimes\mathcal{H}_m$. After post-selecting, $\ket{\psi_d}$ and $\ket{\psi_r}$ represent the successful detection state and rejection state respectively, and $P$ is successful possibility.}
	\label{fig:0}
\end{figure}

\emph{Post-selected measurements and weak-value amplification (WVA).---}%
Post-selected measurements, different from ideal PVMs, have been attracting growing
interest \cite{Kofman2012}. The most notable post-selected detection strategy is  WVA, involving a
complex weak value of the observable and a small number of measurement trials \cite{aharonov1988result,dressel2014colloquium}.
% and post-selection process is performed by a PVM projecting the system on a presented subspace~
%\cite{aharonov1964,hilgevoord2002time}.
%As an effective measurement strategy, the research on it  has been expanding with an increasing rate.
%Moreover, weak value~\cite{aharonov1988result} as one of the most striking developments of it has been widely discussed studied~\cite{hosten2008observation,ritchie1991,dixon2009,pryde2005,pfeifer2011,kim2022}.
For a simple WVA model~\cite{Jozsa2007} as shown in Fig.~\ref{fig:0}(b),
 the ancillary state is initially pre-selected as $\ket{\psi_i}_{\textrm{E}}$, and the sensor state is $\ket{+}_{\textrm{S}}$, where $\ket{\pm}_{\textrm{S}}$ are the eigenstates of $\hat{\sigma}^x$, with $\hat{\sigma}^{x,y,z}$ being Pauli matrices.
The interaction Hamiltonian reads $\hat{H}_{\textrm{SE}}=-\theta\delta(t-t_0)\hat{\sigma}^z_{\textrm{S}}\otimes\hat{A}_{\textrm{E}}$, where $\theta$ is the interaction strength to be estimated, and  $\hbar\equiv1$.
For %a weak interaction
$\theta \to 0$, the evolved joint state is approximately calculated as $\ket{\Psi}_{\textrm{SE}}\simeq\ket{+}_\textrm{S}\otimes\ket{\varphi_i}_\textrm{E}+i\theta\hat{A}_\textrm{E}\ket{-}_\textrm{S}\otimes\ket{\varphi_i}_\textrm{E}$.
By post-selecting the ancilla in the state $\ket{\varphi_f}_\textrm{E}$, the detected sensor state becomes
$\ket{\psi_{{d}}}_\textrm{S}\propto {}_\textrm{E}\langle\varphi_f|\Psi\rangle_{\textrm{SE}} \simeq \ket{+}_\textrm{S}+\exp(i\theta{A}_{\textrm{w}})\ket{-}_\textrm{S}$,
with a success probability of $P_d=|_\textrm{E}\langle\varphi_f|\varphi_i\rangle_\textrm{E}|^2$.
%,
%and the rejected sensor state $\ket{\psi_r}_S\propto {}_E\langle\varphi_f^{\perp }|\Psi\rangle_{SE}$ is neglected.
Here, the weak value ${A}_{\textrm{w}}\equiv{}_\textrm{E}\!\bra{\varphi_f}\hat{A}_\textrm{E}\ket{\varphi_i}_\textrm{E}/{}_\textrm{E}\langle\varphi_f|\varphi_i\rangle_\textrm{E}$ denotes the amplification factor of the signal and grows very large when $P_d\rightarrow 0$.
%the post-selected state  is almost orthogonal to the pre-selected state as ${}_E\langle\psi_f|\psi_i\rangle_E\rightarrow 0$.
Although WVA seems an effective method for improving the SNR in the QPE,
whether the post-selected measurements provide estimation precision superior to conventional strategies remains controversial \cite{brunner2010,starling2009,ferrie2014weak,zhang2015precision,pang2015,jordan2014,knee2016,knee2014amplification,harris2017weak,Tanaka2013,kim2022}.

Generally, a post-selection process can be expressed as a PVM on the joint state $|\Psi\rangle_{\textrm{SE}}$:
$\{\ket{\varphi_f}_\textrm{E}\!\bra{\varphi_f}\otimes\mathbb{I}_\textrm{S},(\mathbb{I}_\textrm{E}-\ket{\varphi_f}_\textrm{E}\!\bra{\varphi_f})\otimes\mathbb{I}_\textrm{S}\}$, where $P_d=|{}_\textrm{E}\langle\varphi_f|\Psi\rangle_{\textrm{SE}}|^2$,
$\ket{\psi_{{d}}}_\textrm{S}\propto {}_\textrm{E}\langle\varphi_f|\Psi\rangle_{\textrm{SE}}$, and $\rho_{r,\textrm{S}}\propto\textrm{Tr}[(\mathbb{I}_\textrm{E}-\ket{\varphi_f}_\textrm{E}\!\bra{\varphi_f})|\Psi\rangle_{\textrm{SE}}\langle\Psi|]$
denote the success probability, detected state, and rejected state, respectively.
The total FI for the post-selection strategy, $F_{\textrm{tot}}$, can be divided into three parts as \cite{zhang2015precision}
\begin{equation}
		F_{\textrm{tot}}[\ket{\Psi}_{\textrm{SE}}]=P_dQ_d[\ket{\psi_d}_\textrm{S}]+P_rQ_r[\rho_{r,\textrm{S}}]+F_\textrm{post}, \label{1}
\end{equation}
with $P_r\equiv1-P_d$ being the rejection probability.
Here, $Q_d$ ($Q_r$) denotes the QFI with respect to $\ket{\psi_d}_\textrm{S}$ ($\rho_{r,\textrm{S}}$), and $P_dQ_d$ ($P_rQ_r$) denotes the effective QFI when considering the success (rejection) probability $P_d$ ($P_r$). The last term $F_\textrm{post}\equiv (\partial_\theta P_d)^2/P_dP_r$ denotes the FI for the post-selection process itself.
Since the POVMs of post-selected measurements may not be optimal for achieving the QFI of the joint state $\ket{\Psi}_{\textrm{SE}}$, %of the combined system $F_Q[\ket{\Psi}_{SE}]$,
we have
 \begin{equation}
	F_{\textrm{tot}}[\ket{\Psi}_{\textrm{SE}}]\le  F_Q[\ket{\Psi}_{\textrm{SE}}], \label{2}
\end{equation}
indicating that the post-selected measurement strategy (including WVA) cannot outperform the optimal conventional strategy~\cite{zhang2015precision,Yang2024}.
Nevertheless, the post-selected strategy can be highly efficient, %and effective against some technical noises,
as $P_dQ_d$ approaches to the total QFI, even
when most of the outcomes are discarded %$P_d\rightarrow0 $
\cite{Tanaka2013,zhang2015precision,pang2015}.

%{\bf Even most of the outcomes are discarded}

%From a perspective of quantum information, the post-selection-strategy measurement is suboptimal with the sensitivity of QPE characterized by QFI.
%Similarly, $P_rQ_r$ denotes the effective QFI of $\ket{\psi_r}_S$.

\emph{Connecting non-Hermitian sensing to post-selected measurements with Naimark dilation---}%
Non-Hermitian systems have been %as a gold mine of new physics have
 attracting growing interest in many fields of frontier physics \cite{ashida2020non,Okuma2023}, among which non-Hermitian sensors are expected to have potential advantages in high-precision sensing~\cite{hokmabadi2019non,yu2020experimental,lai2019observation,hodaei2017enhanced,chen2017exceptional,wiersig2014enhancing,wang2022boosting,liu2016metrology,rosa2021,tang2023,li2023,xu2024single,Ruan2025}.
 Usually, quantum sensing with a non-Hermitian Hamiltonian is implemented
 in the open quantum system with gain and/or loss \cite{Rotter2009}, and thus,
 the environment as a metrological resource cannot be simply neglected,
 see Fig.~\ref{fig:0}(a).
 Here, we apply the Naimark dilation technique \cite{beneduci2020notes} to extend the non-Hermitian
 system into a larger Hermitian one, which is widely used in quantum information theory~\cite{holevo2011probabilistic,holevo2001statistical} and feasible in experiments for simulating a non-Hermitian system~\cite{wu2019observation,yu2022}.

%Naimark dilation as a powerful mathematical tool is widely used in quantum information theory~\cite{beneduci2020notes,holevo2011probabilistic,holevo2001statistical}, which is applied in experiments for simulating non-Hermitian systems~\cite{wu2019observation,yu2022}.
%According to Naimark dilation, a non-unitary evolution generated by non-Hermitian Hamiltonian can be equivalently represented by a unitary one of an enlarged system followed by a PVM, i.e., post-selection process~\cite{gunther2008naimark,huang2019simulating}.

According to the Naimark dilation theorem \cite{gunther2008naimark}, the non-unitary evolution governed by a non-Hermitian Hamiltonian can be represented by a unitary dynamics of an enlarged system followed by a PVM, i.e., a post-selection process~\cite{gunther2008naimark,huang2019simulating}.
For a non-Hermitian Hamiltonian ${H}_\textrm{S}(\theta)$, the evolved state $\ket{\psi(t)}_\textrm{S}$
is decided by the Schr\"{o}dinger equation $i\partial_t \ket{\psi(t)}_\textrm{S}={H}_\textrm{S}(\theta)\ket{\psi(t)}_\textrm{S}$.
The dilated Hermitian Hamiltonian ${H}_{\textrm{SE}}(t)$ should satisfy \cite{huang2019simulating}: $i\partial_t \ket{\Psi(t)}_{\textrm{SE}}={H}_{\textrm{SE}}(t)\ket{\Psi(t)}_{\textrm{SE}}$, where $\ket{\Psi(t)}_{\textrm{SE}}\propto\ket{\psi(t)}_\textrm{S}\otimes\ket{0}_\textrm{E}+\hat{m}(t)\ket{\psi(t)}_\textrm{S}\otimes\ket{1}_\textrm{E}$, $\hat{m}(t)\equiv[\hat{\eta}(t)-\mathbb{I}]^{1/2}$ is a linear operator,
$\hat{\eta}(t)\equiv\mathcal{T}\exp[-i\! \int_{0}^{t}\!d\tau\,{H}_\textrm{S}^\dagger(\tau)]\hat{\eta}_0\overline{\mathcal{T}}\exp[i\! \int_{0}^{t}\!d\tau\,{H}_\textrm{S}(\tau)]$, with $\mathcal{T}$ and
$\overline{\mathcal{T}}$ being the time-ordering and anti-time-ordering operators, respectively, and
$\ket{0,1}$ are eigenstates of $\hat{\sigma}^z$.
The dilated Hamiltonian is written as
 \begin{equation}
		{H}_{\textrm{SE}}(t)={H}^{(1)}_\textrm{S}(t)\otimes\mathbb{I}_\textrm{E}+i{H}^{(2)}_\textrm{S}(t)\otimes\hat{\sigma}^y_\textrm{E}, \label{3}
\end{equation}
where ${H}^{(1)}_\textrm{S}\equiv\{{H}_\textrm{S}+\hat{m}{H}_\textrm{S}\hat{m}+i(\partial_t\hat{m})\hat{m}\}\hat{\eta}^{-1}$, and ${H}^{(2)}_\textrm{S}\equiv\{[{H}_\textrm{S},\hat{m}]-i\partial_t\hat{m}\}\hat{\eta}^{-1}$.
The evolved state of the non-Hermitian system can be obtained from the evolution of the large Hermitian system, followed by post-selecting the environment state in $\ket{0}_\textrm{E}$, written as $\ket{\psi(t)}_\textrm{S}\propto{}_\textrm{E}\langle 0|\Psi(t) \rangle_{\textrm{SE}}$.
%Therefore, a non-Hermitian system can be equivalently represented by a Hermitian evolution with a post-selection process in a extended space twice as large as itself.}

%Thus, quantum sensing with a non-Hermitian Hamiltonian can be
%represented as a post-selected detection strategy
%with a dilated Hermitian Hamiltonian of an extended system, where $|0\rangle_E$
%and $\ket{\psi(t)}_S$ correspond to the post-selected state
%$|\varphi_f\rangle_E$ and the resulting sensor state $|\psi_d\rangle_S$, respectively.
Since $|0\rangle_\textrm{E}$
and $\ket{\psi(t)}_\textrm{S}$ correspond to the post-selected state
$|\varphi_f\rangle_\textrm{E}$ and the resulting sensor state $|\psi_d\rangle_\textrm{S}$ in  WVA, respectively, %a typical post-selected detection protocol,
it is reasonable
that the QFI, $F_Q^{\textrm{nH}}$, for a non-Hermitian (nH)  sensor
can be very large or even tends to infinity, implying an improvement of SNR. However, the environment, interacting with the
non-Hermitian sensor, should be considered as an additional
metrological resource, and
%Due to information loss,
the effective QFI, $P_dF_Q^{\textrm{nH}}$, should be considered when analyzing the sensitivity of non-Hermitian sensors. This approach is similar to neglecting some detection trials in the post-selected detection strategy.
%{\bf Although the QFI $F_Q^{\textrm{nH}}$ for a non-Hermitian sensor can be very large, (1) the environment is not considered (2) similar results are obtained as WVA.}
Using Eqs.~(\ref{1},\ref{2}), we conclude that the
effective QFI for non-Hermitian sensors does not exceed  the QFI
of their dilated Hermitian counterparts as
%
%the evolved state of the
%the joint state is the state of extended system $\ket{\Psi(t)}_{SE}=\mathcal{T}\exp[-i \int_{0}^{t}\!d\tau\,\hat{H}_{SE}(\tau)]\ket{\Psi(0)}_{SE}$ with the post-selection state being $\ket{0}_E$.
%The joint state will be projected on the successful detection state $\ket{\psi_d}_S$ after post-selection, where $\ket{\psi_d}_S$ is equal to the evolved state of the non-Hermitian system $\ket{\psi(t)}_S$.
%Than, the effective QFI of the non-Hermitian sensor is no larger than the QFI of its Hermitian counterpart]
\begin{equation}
%F_{\textrm{nH}}^{\textrm{eff}}=
P_dF_Q^{\textrm{nH}}[\ket{\psi(t)}_{\textrm{S}}] \leq F_Q[\ket{\Psi(t)}_{\textrm{SE}}], \label{4}
\end{equation}
with $P_d=|{}_\textrm{E}\langle 0|\Psi(t) \rangle_{\textrm{SE}}|^2$.
Note that our results also hold for non-Hermitian sensors with %unexpected technical or
experimental imperfections on the detectors that can be expressed as quantum channels, by considering
%the tight upper bound %$\mathcal{C}_Q$
of the noisy QFI \cite{escher2011general,escher2012quantum}.

In addition to the ultimate sensitivity limits, mapping non-Hermitian sensing to a
post-selection QPE strategy offers a novel perspective for comparing the efficiencies of different
types of non-Hermitian sensors. Similar to WVA, the efficiency can be evaluated by comparing the
effective QFI, $P_dF_Q^{\textrm{nH}}$, with the total QFI, $F_Q$, for the dilated Hermitian
system. When the ratio $P_dF_Q^{\textrm{nH}}/F_Q\rightarrow1$ with a small value of $P_d$, it indicates a highly efficient non-Hermitian sensing scheme, in analog with the advantage of ``\emph{when less is more}'' in  WVA \cite{jordan2014}. %
Next, we investigate different types of non-Hermitian sensors from the perspective
of post-selected measurements. %by dilating their non-Hermitian Hamiltonians to Hermitian
%ones in extended Hilbert space.

%{\bf
%Note that this result also holds for the non-Hermitian sensor with
%unexpected decoherence, since xxx.

%In addition,
%xxxx other discussions about the xxx (minimal extension of the system).

%From another view, the non-Hermitian sensor apply the environment (noise)
%to enhance sensing sensitivity.
%(Effectiveness of the non-Hermitian sensing from the perspective of WVA)
%When less is more or when more is less
 %where we have used Eq.~(\ref{1}) and Eq.~(\ref{2}).
%Thus, non-Hermitian sensors are equivalent to post-selection measurements, which are suboptimal even without the consideration of the effect of noise.
%In other words, non-Hermitian sensors lost their potential advantages when considering the resources of environments through Naimark dilation.}

%Later on, we consider several examples.}

\begin{figure}
	\centering
	\includegraphics[width=0.48\textwidth]{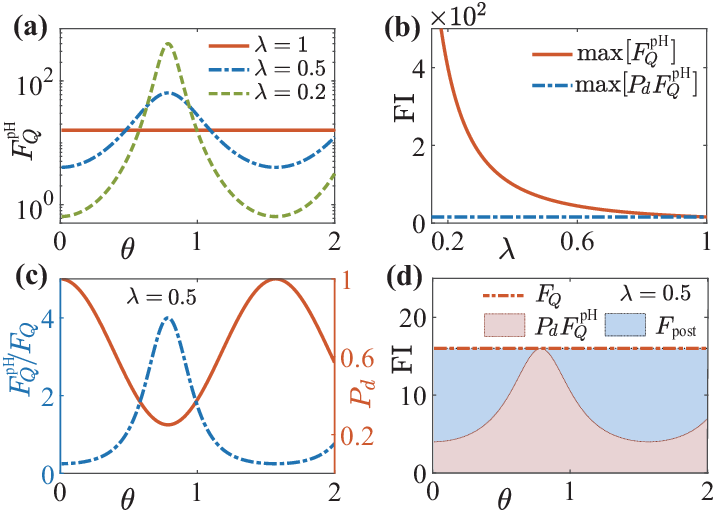}\\
	\caption
	{Pseudo-Hermitian (pH) sensor with the Hamiltonian (\ref{5}).
	(a) QFI, $F_Q^{\textrm{pH}}$, for the sensor state $\ket{\psi(t)}_\textrm{S}$  for $\lambda=1$, $0.5$, $0.2$, when $t=2$.
	(b) The maximum $F_Q^{\textrm{pH}}$ over $\theta$ diverges as $\lambda \to 0$. However, the maximum effective QFI, $P_dF_Q^{\textrm{pH}}$, becomes finite for any value of $\lambda$, with the success probability $P_d$.
	(c) Ratio of the pseudo-Hermitian QFI over the total QFI for the dilated system $F_Q^{\textrm{pH}}/F_Q$ (blue dash-dotted curve) is shown with the success probability $P_d$ (red solid curve) for $\lambda=0.5$.
	(d) The total QFI of the extended system, $F_Q$, the effective QFI,  $P_dQ_d$, and the FI of the post-selection process, $F_{\textrm{post}}$, for $\lambda=0.5$.
	Here, the effective QFI for the rejected state vanishes $P_rQ_r=0$, since $\partial_\theta \ket{\psi_r(t)}_\textrm{S}=0$. }
	\label{fig:1}
\end{figure}

\emph{Pseudo-Hermitian sensor.---}%
We first consider a  pseudo-Hermitian (pH) sensor with a Hamiltonian for $\lambda \in (0,1]$:
\begin{equation}
		{H}_\textrm{S}^{\textrm{pH}}=\theta\begin{pmatrix}~0&~~~\lambda^{-1}\\~\lambda&0\end{pmatrix},\label{5}
\end{equation}
which was believed to enable non-Hermiticity-enhanced sensing~\cite{chu2020quantum,xiao2024non}.
Here, $\theta$ is the unknown parameter to be estimated, and ${H}_\textrm{S}^{\textrm{pH}}$ is non-Hermitian when $\lambda\ne1$.
%($\lambda \in (0,1]$ without loss of generality).
By setting $\ket{0}_\textrm{S}$ as the initial state, the time evolved state under ${H}_\textrm{S}^{\textrm{pH}}$ is $\ket{\psi(t)}_\textrm{S}=[\cos(\theta t)\ket{0}_\textrm{S}-i\lambda \sin(\theta t)\ket{1}_\textrm{S}]/C$,
with $C\equiv[\cos^2(\theta t)+\lambda^2 \sin^2(\theta t)]^{1/2}$, and the QFI with respect to $\ket{\psi(t)}_\textrm{S}$ is calculated as $F_Q^{\textrm{pH}}[\ket{\psi(t)}_\textrm{S}]=4\lambda^2 t^2/C^4$.
The QFI under the non-Hermitian condition ($\lambda\ne1$) is larger than that for the Hermitian case ($\lambda=1$), when choosing a proper parameter range of $\theta$, e.g., see Fig.~\ref{fig:1}(a) for $t=2$.
Moreover, the maximum QFI, $\max_{\theta}\{F_Q^{\textrm{pH}}\}$, diverges as $\lambda \to 0$, see Fig.~\ref{fig:1}(b), since it is proportional to $\lambda^{-2}$, implying
superior performance to the conventional Hermitian case.

%With the Naimark dilation technique,
%a extended system with a post-selected measurement can equivalently represent the dynamics of the pseudo-Hermitian system Eq.~(\ref{5}).
The Naimark-dilated Hermitian Hamiltonian with respect to the pseudo-Hermitian system  (\ref{5}) can be obtained as
\begin{equation}
		{H}_{\textrm{SE}}=\theta\lambda\left(\hat{\sigma}^x_\textrm{S}\otimes\mathbb{I}_\textrm{E}-\sqrt{\lambda^{-2}-1}\hat{\sigma}^y_\textrm{S}\otimes\hat{\sigma}^y_\textrm{E}\right).
	\end{equation}
The time-evolved state of the extended system is $\ket{\Psi(t)}_{\textrm{SE}}=\ket{\psi(t)}_\textrm{S}\otimes\ket{0}_\textrm{E}+i (1-\lambda^{2})^{1/2}\sin(\theta t)\ket{1}_\textrm{S}\otimes\ket{1}_\textrm{E}$, which is initialized at $\ket{\Psi(0)}_{\textrm{SE}}=\ket{0}_\textrm{S}\otimes\ket{0}_\textrm{E}$.
After post-selecting the environment in $\ket{0}_\textrm{E}$ with a success probability of $P_d=[(1-\lambda^2)\sin^2(\theta t)+1]^{-1}$, the time-evolved sensor state $\ket{\psi(t)}_\textrm{S}$ is obtained.
For some values of $\theta$, $F_Q^{\textrm{pH}}[\ket{\psi}_\textrm{S}]/ F_Q[\ket{\Psi}_{\textrm{SE}}]>1$,
see Fig.~\ref{fig:1}(c), where $P_d$ becomes relatively small.
%From the perspective of post-selected measurements,
Here, the effective QFI, $P_dF_Q^{\textrm{pH}}$, should be considered, which does not diverge [see Fig.~\ref{fig:1}(b)].
%, when compared to the QFI $F_Q$ for the extended Hermitian system.
Using Eq.~(\ref{1}), we have $F_Q\ge F_{\textrm{tot}}\ge P_dF_Q^{\textrm{pH}}$, as shown in Fig.~\ref{fig:1}(d), indicating that the pseudo-Hermitian sensor is suboptimal when compared to its dilated Hermitian counterpart.
%This result is also due to the additivity of the FI, since we cannot obtain more FI from a subsystem than from the total system \cite{escher2011general}. %Note that $Q_r=0$, since
%the rejected state does not contain any information about the parameter $\theta$.

Furthermore, Fig.~\ref{fig:1}(d) shows that  $P_dF_Q^{\textrm{pH}} \approx F_Q$, as $\theta\simeq0.785$, indicating that the pseudo-Hermitian sensor can be efficient.
It is because that most information about $\theta$ can be obtained with very few measurement trials on the detected state, which, similar to WVA, could help to
overcome some technical noise \cite{jordan2014,harris2017weak,xiao2024non}. In addition, since the rejected state does not
contain information about $\theta$, it is straightforward  that $Q_r=0$.

\emph{Two types of EP-based non-Hermitian sensors.---}%
In non-Hermitian systems with EPs, where gain and loss can be
perfectly balanced, exotic behaviors are predicted to occur \cite{Miri2019,Okuma2023,bender2024}  with promising applications, e.g., EP-based sensing \cite{wiersig2020review,wang2020petermann,lai2019observation,yu2020experimental}.
Here, we consider two types of EP-based sensors
%, which provide another road map for quantum sensing.
%We consider two kinds of EP-based sensors
without and with $\mathcal{PT}$-symmetry, respectively.

\begin{figure}
	\centering
	\includegraphics[width=0.48\textwidth]{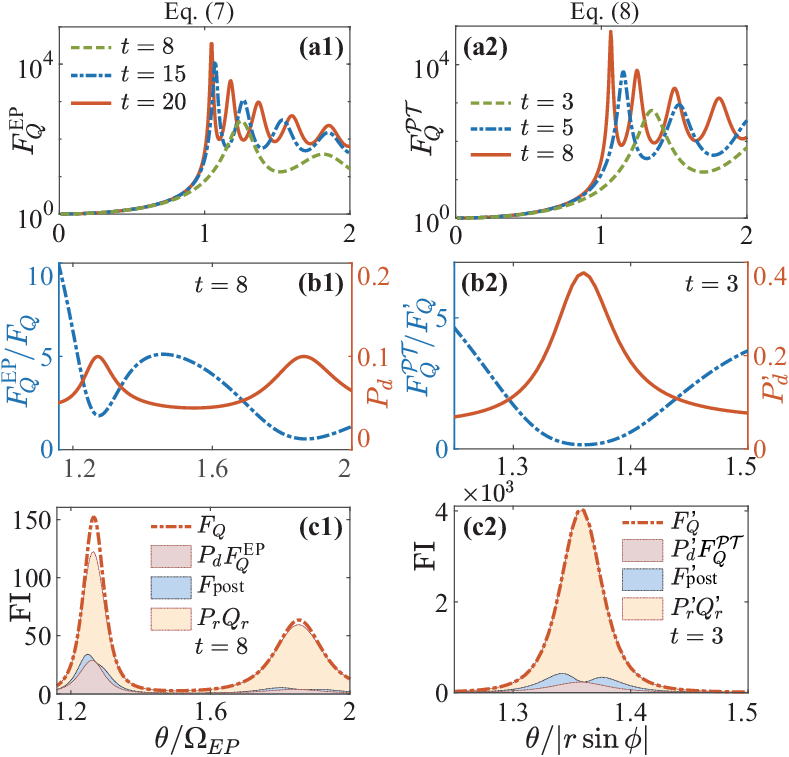}\\
	\caption
	{Two EP-based sensors with Hamiltonians in Eqs.~(\ref{6}) and (\ref{7}), respectively.
		(a1) QFI, $F_Q^{\textrm{EP}}$,  for the EP-based sensor (\ref{6}) for evolution times $t=8$, $15$, $20$, when choosing $\Omega_{\textrm{EP}}=\omega_{\textrm{ccw}}$.
		(a2) QFI,  $F_Q^{\mathcal{PT}}$, for the EP-based sensor (\ref{7}) for $t=3$, $5$, $8$, with $\phi=\pi/4$.
		(b1) and (b2) Ratios of the QFI for the EP-based sensors $F_Q^{\textrm{EP}}$ and $F_Q^{\mathcal{PT}}$ to the total QFI $F_Q$ and $F'_{Q}$ for the extended systems (blue dashed-dotted curves) are shown with the corresponding success probabilities $P_d$ and $P'_d$, respectively.
		(c1) and (c2) The total effective QFI, the effective QFI of EP-based sensors $P_dF_Q^{\textrm{EP}}$ and $P'_dF_Q^{\mathcal{PT}}$, the effective QFI for the rejected states $P'_rQ'_r$ and $P'_rQ'_r$, and the FI for post-selection process $F_{\textrm{post}}$ and $F'_{\textrm{post}}$ are shown for $t=8$ and $3$, respectively.
}
	\label{fig:2}
\end{figure}

The first model has a non-Hermitian Hamiltonian:
\begin{equation}
		\hat{H}_\textrm{S}^{\textrm{EP}}=\begin{pmatrix}\omega_{\textrm{cw}}&i\Omega_{\textrm{EP}}/2\\i\Omega_{\textrm{EP}}/2&\omega_{\textrm{ccw}}\end{pmatrix},\label{6}
\end{equation}
which has been experimentally realized in a Brillouin ring laser gyroscope~\cite{wang2020petermann,lai2019observation}.
This EP-based sensor estimates the frequency difference $\theta\equiv\omega_{\textrm{cw}}-\omega_{\textrm{ccw}}$ by measuring the eigenenergy difference $\Delta E$ of the Hamiltonian in Eq.~(\ref{6}).
The eigenenergies of $\hat{H}_\textrm{S}^{\textrm{EP}}$ are $E_\pm\equiv[\omega_{\textrm{cw}}+\omega_{\textrm{ccw}}\pm (\theta^2-\Omega_{\textrm{\textrm{EP}}}^2)^{1/2}]/2$, with $\Delta E=(\theta^2-\Omega_{\textrm{\textrm{EP}}}^2)^{1/2}$, which vanishes at two EPs ($\theta=\pm\Omega_{\textrm{EP}}$).
The estimation sensitivity $\delta \theta$ is proportional to the inverse differential $(\partial\Delta E/\partial \theta)^{-1}$, and the differential diverges at EPs, implying highly sensitive estimation in absence of noise~\cite{lai2019observation}.
For the initial state $\ket{0}_\textrm{S}$, as time becomes sufficiently long, the QFI diverges near the EPs, see Fig.~\ref{fig:2}(a).
Then, we consider the dilated Hamiltonian $\hat{H}^{\textrm{EP}}_{\textrm{SE}}(t)$ using Eq.~(\ref{3}), see Supplementary Materials (SM) \cite {SM} for more details.
For the initial state $\ket{\Psi(0)}_{\textrm{SE}}=\ket{0}_\textrm{S}\otimes\ket{0}_\textrm{E}+[\hat{\eta}(0)-\mathbb{I}]^{1/2}\ket{0}_\textrm{S}\otimes\ket{1}_\textrm{E}$, the time-evolved state
 satisfies $i\partial_t \ket{\Psi(t)}^{\textrm{EP}}_{\textrm{SE}}=\hat{H}^{\textrm{EP}}_{\textrm{SE}}(t)\ket{\Psi(t)}_{\textrm{SE}}$.
 % with the initial state being $\ket{\Psi(0)}_{SE}=\ket{0}_S\otimes\ket{0}_E+[\hat{\eta}(t=0)-\mathbb{I}]^{1/2}\ket{0}_S\otimes\ket{1}_E$.
Without loss of generality, we set $\hat{\eta}(0)=100$ for $t \le 15$ and calculate the QFI for the EP-based and the extended Hermitian sensors, respectively, which are compared to the success probability $P_d$ in Fig.~\ref{fig:2}(b1).
Figure~\ref{fig:2}(c1) shows that the effective QFI for this EP-based sensor is much smaller than the QFI for the dilated Hermitian system as $P_dF_Q^{\textrm{EP}}\leq F_Q$, which complies with Eq.~(\ref{4}). It demonstrates that the EP-based sensor (\ref{6}) cannot outperform the conventional Hermitian sensor.

Next, we consider a two-level $\mathcal{PT}$-symmetric system \cite{yu2020experimental} with a Hamiltonian:
\begin{equation}
	\begin{split}
		\hat{H}_{\textrm{S}}^{\mathcal{PT}}=\begin{pmatrix}re^{i\phi}&\theta\\\theta&re^{-i\phi}\end{pmatrix},\label{7}
	\end{split}
\end{equation}
where $\theta$ is the unknown parameter to be estimated.
%Equation~(\ref{7}) has $\mathcal{PT}$-symmetric, as $[\hat{H}_{\mathcal{PT}},\hat{P}\hat{T}]=0$, with $\hat{P}$ and $\hat{T}$ being the parity and the time-reversal operators respectively.
Its eigenvalues are $E_{\pm}=r\cos\phi\pm(\theta^2-r^2\sin^2\phi)^{1/2}$, and this $\mathcal{PT}$-symmetric system has two EPs ($\theta=\pm r\sin\phi$).
When $|\theta|>|r\sin\phi|$, the eigenvalues are real;
%with symmetry being kept,
otherwise, the  eigenvalues are complex since the $\mathcal{PT}$-symmetry is broken~\cite{ozdemir2019parity,el-ganainy2018,bender2024}.
Similarly, a highly sensitive sensor is theoretically predicted near the EPs, due to the divergence of the inverse differential $(\partial\Delta E/\partial \theta)^{-1}$, where
the QFI can be arbitrary large for $t \rightarrow \infty$, see Fig.~\ref{fig:2}(a2).
Considering the post-selection on the dilated Hermitian system, the effective QFI of the $\mathcal{PT}$-symmetric sensor  is also smaller than the total QFI as $P'_dF_Q^{\mathcal{PT}}\leq F'_Q$, see Fig.~\ref{fig:2}(c2).

Therefore, by relating EP-based sensing to post-selected measurements, both EP-based sensors
cannot outperform their extended Hermitian counterparts even with divergent QFI near EPs.
Moreover, the maximum effective QFI appear when the success probabilities $P_d$ and
$P'_d$ achieve their local maxima, see Fig.~\ref{fig:2}(b1,b2).
In the context of post-selected measurements, it corresponds to an inefficient
post-selection strategy, since ``\emph{when more is less}''.
In comparison, the effective QFI for the pseudo-Hermitian sensor achieves the maximum when the success probability
achieves its local minimum [Fig.~\ref{fig:2}(c2)], i.e., ``\emph{when less is more}''. Another  EP-based sensing scheme \cite{Ruan2025}, where the dynamics of a loss-loss system is mathematically equivalent to that of a gain-loss system apart from a global exponential decay, is discussed in the SM \cite{SM}.

%Therefore, we have demonstrated with two examples that the EP-based sensors are suboptimal compared to their Hermitian counterparts even at EPs.
%In addition, different from the example of pseudo-Hermitian sensor and WVA strategy, these two EP-based sensors have loss vast quantities of information during post-selection process, which means the strategy here is even more ineffective.
%From a perspective of post-selected measurements by considering a extended Hermitian system, we show the non-Hermitian sensors are suboptimal in full consideration of the cost of additional resources in their environment.

\emph{Conclusions and discussions.---}%
In summary,  by employing the Naimark dilation method, we establish a connection between non-Hermitian quantum sensing
and a post-selection process implemented on an extended Hermitian system.
Through analyzing the effective QFI of the non-Hermitian sensing from the perspective of post-selected measurements, we demonstrate that non-Hermitian sensors  exhibit suboptimal performance compared to their extended Hermitian counterparts, when all information is harnessed.
Analogous to WVA, the efficiency of non-Hermitian quantum sensing, quantified by the ratio of the effective QFI to the total QFI,
can be optimized under post-selected measurement protocols with minimal experimental trials.
Our work establishes an alternative framework for understanding non-Hermitian sensing from the perspective of post-selected measurements and facilitates the design of robust quantum metrological protocols against technical noise.

Note that our results based on the conservation of total QFI do not mean that non-Hermitian sensing is completely ineffective. Several implementations exhibit enhanced parameter sensitivity under specific conditions \cite{Ruan2025,xiao2024non}.
First, since full control of the sensor-environment system and the extraction of all information from the environment space is experimentally infeasible, non-Hermitian sensing provides a practical method to harness useful quantum metrological resources from the interaction with the environment.
In addition, the Naimark dilation method may not be the minimal extension of
the system \cite{Naghiloo2019}, but fortunately, in most experiments, the environmental dimension is much larger than the Naimark-dilated ancillary system dimension.
Then, the efficiency of non-Hermitian sensors can be evaluated using our framework. For instance, the pseudo-Hermitian sensor (\ref{5}) relates to the post-selection protocol that requires very few trials of measurements and has an effective QFI that equals the total QFI, showing potential advantages when the control of the environment is limited.
Since post-selection detection protocols can suppress some forms of technical noise \cite{knee2014amplification,harris2017weak,Peng2014},
our theoretical framework will further motivate the design of practical noise-resilient quantum metrology that leverages the interaction with the environment
as a resource rather than a limitation.
Further research on this topic would include the use of the
quantum correlation measurement \cite{Wang2021,Shen2023} to remove the classical noise
and practical QPE with different statistical methods, e.g,  the maximum likelihood analysis \cite{Bobroff1986} and Bayesian analysis \cite{Tsang2012}.
%When the sensing resources is limited, non-Hermitian sensors lost its advantage.
%lthough the resources are always limited in practical sensing scenarios, it is difficult to use all the resources.
%Since post-selection protocols can suppress some technical noises, thereby guiding the design of noise-resilient quantum sensing architectures.
%Non-Hermitian sensors provide a method to use partial environmental resources which is hard to be used, in this sense, they can be effective in actual sensing technology.

\begin{acknowledgments}
We would like to thank Qing Ai %Keyu Xia
for useful discussions.
T.L. acknowledges the support from the National Natural Science Foundation of China (Grant No.~12274142), the
Key Program of the National Natural Science Foundation of China (Grant No.~62434009),  the Fundamental
Research Funds for the Central Universities (Grant No.~2023ZYGXZR020), the Introduced Innovative Team
Project of Guangdong Pearl River Talents Program (Grant No.~2021ZT09Z109), and the Startup Grant
of South China University of Technology (Grant No.~20210012).
K.X. is partly supported by the National Natural Science
Foundation of China (Grant No.~92365107),
the National Key R\&D Program of China (Grant No.~2019YFA0308700), the Program for Innovative Talents and
Teams in Jiangsu (Grant No.~JSSCTD202138), and the National University of Defense Technology Independent Innovation Science Foundation (Grant No.~24-ZZCX-JDZ-11).
Y.R.Z. is supported in part by:
the National Natural Science Foundation of China (Grant No.~12475017), the Natural Science Foundation of
Guangdong Province (Grant No.~2024A1515010398), and the Startup Grant of South China University of
Technology (Grant No.~20240061).
F.N. is supported in part by:
%the Nippon Telegraph and Telephone Corporation (NTT) Research,
the Japan Science and Technology Agency (JST) [via the CREST Quantum Frontiers
program Grant No.~JPMJCR24I2, the Quantum Leap Flagship Program (Q-LEAP), and the Moonshot R\&D
Grant No.~JPMJMS2061].
%, and the Office of Naval Research (ONR) Global (Grant No.~N62909-23-1-2074).
%Y.R.Z.
%This work is partially supported by the National Natural Science Foundation of China (Grant No.~12475017) and the
%Natural Science Foundation of Guangdong Province (Grant No.~2024A1515010398).
\end{acknowledgments}

\bibliography{nonhermitian.bib}

\end{document}

\title{SUPPLEMENTAL MATERIAL:\\Non-Hermitian sensing from the perspective of post-selected measurements}
 \author{Neng Zeng}
 \affiliation{School of Physics and Optoelectronics, South China University of Technology, Guangzhou 510640, China}

 \author{Tao Liu}
 \affiliation{School of Physics and Optoelectronics, South China University of Technology, Guangzhou 510640, China}

 \author{Keyu Xia}
 \affiliation{College of Engineering and Applied Sciences, National Laboratory of Solid State Microstructures, Nanjing University, Nanjing 210023, China}
 \affiliation{Shishan Laboratory, Suzhou Campus of Nanjing University, Suzhou 215000, China}

 \author{Yu-Ran Zhang}
 \email{yuranzhang@scut.edu.cn}
 \affiliation{School of Physics and Optoelectronics, South China University of Technology, Guangzhou 510640, China}

 \author{Franco Nori}
 %\affiliation{Theoretical Quantum Physics Laboratory, Cluster for Pioneering Research, Wako-shi, Saitama 351-0198, Japan}
 \affiliation{Center for Quantum Computing, RIKEN, Wakoshi, Saitama 351-0198, Japan}
 \affiliation{Physics Department, University of Michigan, Ann Arbor, Michigan 48109-1040, USA}

 \date{\today}% It is always \today, today,
 %\pacs{}

 \maketitle

 %\clearpage

 %\appendix
 \renewcommand{\theequation}{S\arabic{equation}}
 %redefine the command that creates the equation no.
 \setcounter{equation}{0}  % reset counter

 \renewcommand{\thefigure}{S\arabic{figure}}
 %redefine the command that creates the equation no.
 \setcounter{figure}{0}  % reset counter

 \renewcommand{\thetable}{S\arabic{table}}
 %redefine the command that creates the equation no.
 \setcounter{table}{0}

 %\renewcommand{\thesection}{S\arabic{section}}
 %redefine the command that creates the equation no.
 %\setcounter{section}{0}

 \section{Total Fisher information for the post-selection process}
 Here we analyze the total Fisher information for the post-selection process.
 For the post-selected measurement on the joint state $\ket{\Psi}_{\textrm{SE}}$,
 the success probabilities of obtaining the detected state $\ket{\psi_d}_\textrm{S}$ and the rejected state $\rho^{r}_{\textrm{S}}$ are $P_d$ and $P_r\equiv1-P_d$, respectively.
 Subsequent measurements on the sensor state yield outcomes $x$
 with probabilities $p(x|d)$ and $p(x|r)$ for the detected and rejected states, respectively.
 The total Fisher information for the post-selection process can be calculated as
 \begin{align}
     &F_{\textrm{tot}}=\int_{}^{}\!\!dx\,{\frac{[\partial{_\theta}P_dp(x|d)]^2}{P_dp(x|d)}}+\int_{}^{}\!\!dx\,{\frac{[\partial{_\theta}P_rp(x|r)]^2}{P_rp(x|r)}}\nonumber\\
     &=P_d\!\!\int_{}^{}\!\!dx\,{\frac{[\partial{_\theta}p(x|d)]^2}{p(x|d)}}+P_r\!\!\int_{}^{}\!\!dx\,{\frac{[\partial{_\theta}p(x|r)]^2}{p(x|r)}}
     +\!\!\sum_{i=d,r}\!\!\frac{(\partial_\theta P_i)^2}{P_i}\nonumber\\
     &=P_dF_d+P_rF_r+F_\textrm{post} \textrm{.}
 \end{align}
 Here, $F_d$ and $F_r$ represent the Fisher information for $\ket{\psi_d}_\textrm{S}$ and $\rho^{r}_{_\textrm{S}}$, respectively, and $F_\textrm{post}$ denotes the Fisher information for the post-selection process itself.

 In the ideal case, where the POVMs $\{\hat{E}_d\}$ and $\{\hat{E}_r\}$ performed on the the detected and rejected states, respectively, are assumed to be optimal, the total Fisher information $F_{\textrm{tot}}$ for the post-selection process can be optimized as~\cite{zhang2015precision}
 \begin{align}
     F_{\textrm{tot}}^\text{opt}[\ket{\Psi}_{\textrm{SE}}]&\equiv\max_{\{\hat{E}_{d}\}}\max_{\{\hat{E}_{r}\}} F_{\textrm{tot}}\\
     &=P_dQ_d[\ket{\psi_d}_{\textrm{S}}]+P_rQ_r[\rho_{r,\textrm{S}}]+F_p,
 \end{align}
 where $Q_d$ and $Q_r$ denote the quantum Fisher information of $\ket{\psi_d}_\textrm{S}$ and $\rho_{r,\textrm{S}}$, respectively.

 \section{Naimark dilation of non-Hermitian Hamiltonians}\label{II}
 Under the impact of a non-Hermitian Hamiltonian $H_\textrm{S}(t)$, the time-evolved state $\ket{\psi(t)}_\textrm{S}$ is governed by the Schr\"{o}dinger equation with $\hbar\equiv1$ as
 \begin{equation}
     i\partial_t\ket{\psi(t)}_\textrm{S}=H_\textrm{S}(t)\ket{\psi(t)}_\textrm{S}.\label{0}
 \end{equation}
 According to the Naimark dilation theorem \cite{dalla2019naimark,beneduci2020notes},
 this non-unitary dynamics can be equivalently represented as the dynamics of an extended system-environment Hermitian system followed by post-selection.
 The time evolution with the dilated Hermitian Hamiltonian $H_{\textrm{SE}}(t)$ satisfies
 \begin{equation}
     i\partial_t\ket{\Psi(t)}_{\textrm{SE}}=H_{\textrm{SE}}(t)\ket{\Psi(t)}_{\textrm{SE}}.\label{1}
 \end{equation}
The joint state $\ket{\Psi(t)}_{\textrm{SE}}$ evolves as  \begin{equation}
 \ket{\Psi(t)}_{\textrm{SE}}\propto\ket{\psi(t)}_\textrm{S}\otimes\ket{0}_\textrm{E}+\hat{m}(t)\ket{\psi(t)}_\textrm{S}\otimes\ket{1}_\textrm{E}.
  \end{equation}
  where $\hat{m}(t)$ is a linear time-dependent operator, $\ket{0}$ and $\ket{1}$ are eigenstates of $\sigma^z$, with $\hat{\sigma}^{x,y,z}$ being Pauli matrices.

  Without loss of generality, $H_{\textrm{SE}}(t)$ can be written as
 \begin{equation}
     H_{\textrm{SE}}(t)=\left(\begin{array}{c|c}H^{(1)}_\textrm{S}(t)&H^{(2)}_\textrm{S}(t)\\\hline H^{{(2)}\dagger}_\textrm{S}(t)&H^{(4)}_\textrm{S}(t)\end{array}\right),\label{2}
 \end{equation}
 where $H^{(1)}_\textrm{S}(t)$ and $H^{(4)}_\textrm{S}(t)$ are Hermitian. By substituting Eqs.~(\ref{0},\ref{2}) into Eq.~(\ref{1}), we have
 \begin{align}
   \begin{split}
     &H^{(1)}_\textrm{S}(t)+H^{(2)}_\textrm{S}(t)\hat{m}(t)-H_\textrm{S}(t)=0\label{3},\\
   \end{split}\\
   \begin{split}
     &H^{{(2)}\dagger}_\textrm{S}(t)+H^{(4)}_\textrm{S}(t)\hat{m}(t)-\hat{m}(t)H_\textrm{S}(t)-i\partial_t\hat{m}(t)=0.\label{4}
   \end{split}
 \end{align}
 The Hermitian conjugate of these equations gives:
 \begin{align}
   \begin{split}
     &H^{(1)}_\textrm{S}(t)+\hat{m}(t)^\dagger H^{{(2)}}_\textrm{S}(t)^\dagger-H_\textrm{S}(t)^\dagger=0\label{5},\\
   \end{split}\\
   \begin{split}
     &H^{(2)}_\textrm{S}(t)+\hat{m}(t)^\dagger H^{(4)}_\textrm{S}(t)-H_\textrm{S}(t)^\dagger\hat{m}(t)^\dagger+i\partial_t\hat{m}(t)^\dagger=0\label{6}.
   \end{split}
 \end{align}
The unitary time evolution of the extended Hermitian system requires:
 \begin{align}
     {\partial_t}~{_{\textrm{SE}}\langle \Psi(t)|\Psi(t)\rangle_{\textrm{SE}}}=0,\label{7}
   \end{align}
which can be calculated as
 \begin{align}
     0&=\frac{\partial}{\partial t}{_\textrm{S}\!\bra{\psi(t)}}[\hat{m}(t)^\dagger\hat{m}(t)+{\mathbb{I}}]\ket{\psi(t)}_\textrm{S}\nonumber\\
     &={_\textrm{S}\!\bra{\psi(t)}}i[H_\textrm{S}(t)^\dagger\hat{\eta}(t)-\hat{\eta}(t)H_\textrm{S}(t)-i{\partial_t \hat{\eta}(t)}]\ket{\psi(t)}_\textrm{S},\nonumber
 \end{align}
 where we have defined $\hat{\eta}(t)\equiv\hat{m}^\dagger(t)\hat{m}(t)+\hat{\mathbb{I}}$.
We further have
 \begin{equation}
 H_\textrm{S}^\dagger(t)\hat{\eta}(t)-\hat{\eta}(t)H_\textrm{S}(t)-i{\partial_t \hat{\eta}(t)}=0,\label{8}
 \end{equation}
and the solution takes the form:
 \begin{equation}
     \hat{\eta}(t)=\mathcal{T}e^{-i \int_{0}^{t}\!d\tau H_\textrm{S}(\tau)^\dagger}\hat{\eta}(0)\overline{\mathcal{T}}e^{i \int_{0}^{t}\!d\tau\,H_\textrm{S}(\tau)}, \label{9}
 \end{equation}
 with $\mathcal{T}$ and $\overline{\mathcal{T}}$  being time-ordering operator and anti-time-ordering operators, respectively. Note that the form of $\hat{\eta}(0)$ is indeterminate.
 The operator $\hat{m}(t)$ is given by:
 \begin{equation}
     \hat{m}(t)=\hat{U}[\hat{\eta}(t)-{\mathbb{I}}]^{\frac{1}{2}},\label{00}
 \end{equation}
 where $\hat{U}$ is an arbitrary unitary operator.
 %(we can always let $\hat{m}(t)$ is existent by selecting appropriate $\hat{\eta}_0$, see $A_3$ for more details ).
 For simplicity, we set $\hat{U}={\mathbb{I}}$, and Eq.~(\ref{00}) becomes  \begin{equation}\hat{m}(t)=[\hat{\eta}(t)-{\mathbb{I}}]^{\frac{1}{2}}.
  \end{equation}
In addition, it is required to choose an appropriate Hermitian $\hat{\eta}(0)$ to
keep $[\hat{\eta}(t)-{\mathbb{I}}]$ a positive operator, see Sec.~\ref{III} for details, such
that $\hat{m}(t)$ is Hermitian.
 By using Eqs.~(\ref{3}--\ref{6}), one possible solution for Eq.~(\ref{2}) can be obtained as \cite{wu2019observation}
 \begin{align}
     &H^{(1)}_\textrm{S}(t)=H^{(4)}_\textrm{S}(t)=\{H_\textrm{S}+\hat{m}H_\textrm{S}\hat{m}+i{(\partial_t \hat{m})}\hat{m}\}\hat{\eta}^{-1},\nonumber\\
     &H^{(2)}_\textrm{S}(t)=-H^{(2)\dagger}_\textrm{S}(t)=\{[H_\textrm{S},\hat{m}]-i{\partial_t \hat{m}}\}\hat{\eta}^{-1}.\label{10}
 \end{align}
 After some calculations, the extended Hamiltonian can be written as
 \begin{equation}
     H_{\textrm{SE}}(t)=H^{(1)}_\textrm{S}(t)\otimes{\mathbb{I}}_\textrm{E}+iH^{(2)}_\textrm{S}(t)\otimes\hat{\sigma}^y_\textrm{E}.\label{15}
 \end{equation}

Next,
 $H$ called as a $\zeta$-pseudo-Hermitian Hamiltonian \emph{iff} there exists a Hermitian operator $\hat{\zeta}$, satisfying %\cite{mostafazadeh2010}
 \begin{equation}
 \hat{\zeta}H^\dagger=H\hat{\zeta}.
\end{equation}
  Assuming that $H_\textrm{S}$ is a time-independent $\zeta$-pseudo-Hermitian Hamiltonian, we can simplify the Naimark dilation by choosing a positive $\hat{\zeta}$.
 We let $\hat{\eta(0)}=\hat{\zeta}/\nu_\zeta$, where $\nu_\zeta$ is the minimum eigenvalue of $\hat{\zeta}$.
 Here, $\hat{\eta}$ can be obtained with Eq.~(\ref{9}) as
 \begin{equation}
   \begin{split}
     \hat{\eta}(t)=e^{-iH_\textrm{S}^\dagger t}\hat{\eta}(0)e^{iH_\textrm{S}t}=	\hat{\eta}(0).
   \end{split}
 \end{equation}
 Here, $\hat{\eta}=\hat{\zeta}/\nu_\zeta$ and 	$\hat{m}=[\hat{\eta}-{\mathbb{I}}]^{1/2}$ are both time-independent in this case. Thus,  Eq.~(\ref{10}) can be simplified as
 \begin{align}
     &H^{(1)}_\textrm{S}=H^{(4)}_\textrm{S}=(H_\textrm{S}+\hat{m}H_\textrm{S}\hat{m})\hat{\eta}^{-1},\nonumber\\
     &H^{(2)}_\textrm{S}=-H^{(2)\dagger}_\textrm{S}=[H_\textrm{S},\hat{m}]\hat{\eta}^{-1},\label{17}
 \end{align}
 where the extended Hamiltonian $H_{\textrm{SE}}$ expressed in Eq.~(\ref{15}) is also time-independent.

 \section{Universality of Naimark dilation method}\label{III}
 To ensure $\hat{m}(t)=[\hat{\eta}(t)-{\mathbb{I}}]^{\frac{1}{2}}$ to be Hermitian, we need to choose appropriate $\hat{\eta}(0)$ to keep $\hat{\eta}(t)-{\mathbb{I}}$ positive.
  First, we choose an arbitrary positive operator
\begin{equation}
  \hat{\eta}^{\prime}(0)=\hat{\gamma}^\dagger\hat{\gamma}.
  \end{equation}
  By using Eq.~(\ref{9}), we have
 \begin{equation}
     \hat{\eta}^{\prime}(t)=[\hat{\gamma}\overline{\mathcal{T}}e^{i \int_{0}^{t}\!d\tau\,H_\textrm{S}(\tau)}]^\dagger[\hat{\gamma}\overline{\mathcal{T}}e^{i \int_{0}^{t}\!d\tau\,H_\textrm{S}(\tau)}],
 \end{equation}
 which is still positive.
 By supposing the minimum eigenvalue of $\hat{\eta}^{\prime}(t)$ is $\nu^{\prime}$ and defining
 \begin{equation}
     \hat{\eta}(0)\equiv\frac{\hat{\eta}^{\prime}(0)}{\nu^{\prime}},
 \end{equation}
we ensure that
  \begin{equation}\hat{\eta}(t)-{\mathbb{I}}=\mathcal{T}e^{-i \int_{0}^{t}\!d\tau\,H_\textrm{S}(\tau)^\dagger}\hat{\eta}(0)\overline{\mathcal{T}}e^{i \int_{0}^{t}\!d\tau\,H_\textrm{S}(\tau)}-{\mathbb{I}}
    \end{equation}
    is positive, for the saturation of a Hermitian $\hat{m}(t)$.

\section{Dilation of Pseudo-Hermitian Hamiltonian}
 The pseudo-Hermitian Hamiltonian is written as
 \begin{equation}
     H_{\textrm{S}}^{\textrm{pH}}=\theta\begin{pmatrix}0&\lambda^{-1}\\\lambda&0\end{pmatrix},
 \end{equation}
 where $\lambda \in (0,1]$, and $\theta$ is the unknown parameter to be estimated.
 The Hamiltonian $H_{\textrm{S}}^{\textrm{pH}}$ is non-Hermitian when $\lambda \ne 1$.
 By setting $\ket{0}_\textrm{S}$ as the initial state, the time-evolved state under the
 impact of $H_{\textrm{S}}^{\textrm{pH}}$ is calculated as
 \begin{align}
     \ket{\psi(t)}_\textrm{S}&=\frac{\exp(-iH_{\textrm{S}}^{\textrm{pH}}t)\ket{0}_\textrm{S}}{\textrm{tr}[\bra{\psi(0)}\exp(iH_{\textrm{S}}^{\textrm{pH}\dag} t)\exp(-iH_{\textrm{S}}^{\textrm{pH}}t)\ket{\psi(0)}]^{1/2}}\nonumber\\
     &=[\cos(\theta t)\ket{0}_\textrm{S}-i\lambda \sin(\theta t)\ket{1}_\textrm{S}]/C,
 \end{align}
 with $C\equiv{[\cos^2(\theta t)+\lambda^2 \sin^2(\theta t)]^{1/2}}$ is the normalization coefficient.
 The quantum Fisher information for a pure state $\ket{\psi_\theta}$ can be expressed as $F_Q(\theta)=4(\langle  \partial_\theta\psi_\theta|\partial_\theta\psi_\theta\rangle+|\langle\psi_\theta|\partial_\theta\psi_\theta\rangle|^2)$~\cite{giovannetti2011advances,giovannetti2006quantum}.
 We calculate the quantum Fisher information for $\ket{\psi(t)}_\textrm{S}$ as
 \begin{equation}
   F_Q[\ket{\psi(t)}_\textrm{S}]=\frac{4\lambda^2 t^2}{C^4}.
 \end{equation}

  With the Naimark dilation method as discussed in Sec.~\ref{II}, a unitary dynamics of an enlarged system can be used to equivalently represent the dynamics of this pseudo-Hermitian system.
  Thus, we can construct an appropriate extended system by using the time-independent dilation method, see Sec.~\ref{II} for more details.
  Here, $H_{\textrm{S}}^{\textrm{pH}}$ is a $\eta$-pseudo-Hermitian Hamiltonian, satisfying  \begin{equation}
  H_{\textrm{S}}^{\textrm{pH}\dagger}\hat{\eta}=\hat{\eta}H_{\textrm{S}}^{\textrm{pH}}.
   \end{equation}
  The Hermitian operator $\hat{\eta}$ is calculated as
   \begin{equation}
   \hat{\eta}=\begin{pmatrix}1&a\\a&\lambda^{-2}\end{pmatrix},
  \end{equation}
  with $a$ being an arbitrary real number.
  Without loss of generality, we set $a=0$, and the operator $\hat{\eta}$ is given by:
   \begin{equation}
   \hat{\eta}=\begin{pmatrix}1&0\\0&\lambda^{-2}\end{pmatrix}.
  \end{equation}
  Here, $(\hat{\eta}-{\mathbb{I}})$ is positive.
  With Eq.~(\ref{00}), $\hat{m}$ can be written as
   \begin{equation}
   \hat{m}=\begin{pmatrix}0&0\\0&\sqrt{\lambda^{-2}-1}\\\end{pmatrix}.
  \end{equation}
  With Eq.~(\ref{17}), $H^{(1)}_\textrm{S}$ and $H^{(2)}_\textrm{S}$ are calculated as
  \begin{align}
   \begin{split}
   &H^{(1)}_\textrm{S}=\theta\lambda\hat{\sigma}^x_\textrm{S},\\
   &H^{(2)}_\textrm{S}=i\theta\lambda\sqrt{\lambda^{-2}-1}\hat{\sigma}^y_\textrm{S}.\label{26}
   \end{split}
  \end{align}
 Taking Eq.~(\ref{26}) into Eq.~(\ref{15}), the dilated Hamiltonian reads
   \begin{equation}
 H_{\textrm{SE}}=\theta\lambda(\hat{\sigma}^x_\textrm{S}\otimes\hat{\mathbb{I}}_\textrm{E}-\sqrt{\lambda^{-2}-1}\hat{\sigma}^y_\textrm{S}\otimes\hat{\sigma}^y_\textrm{E}).
  \end{equation}
  The initial extended state is chosen as
  \begin{equation}
     \begin{split}
   \ket{\Psi(0)}_{\textrm{SE}}&\propto\ket{0}_\textrm{S}\otimes\ket{0}_\textrm{E}+\hat{m}\ket{0}_\textrm{S}\otimes\ket{1}_\textrm{E}\\
     &=\ket{0}_\textrm{S}\otimes\ket{0}_\textrm{E},
  \end{split}
  \end{equation}
 and the time evolution of the extended system is governed by $\hat{H}_{\textrm{SE}}$ as
  \begin{align}
   \ket{\Psi(t)}_{\textrm{SE}}&=e^{-i\hat{H}_{\textrm{SE}}t}\ket{\Psi(0)}_{\textrm{SE}}\\
   &=\ket{\psi(t)}_\textrm{S}\otimes\ket{0}_\textrm{E}+\hat{m}\ket{\psi(t)}_\textrm{S}\otimes\ket{1}_\textrm{E}\\
   &=\ket{\psi(t)}_\textrm{S}\otimes\ket{0}_\textrm{E}+i\sqrt{1-\lambda^{2}}\sin(\theta t)\ket{1}_\textrm{S}\otimes\ket{1}_\textrm{E}.
\end{align}
  Therefore, by post-selecting the corresponding environment state in $\ket{0}_\textrm{E}$,  the evolved state of the pseudo-Hermitian system can be equivalently obtained as
  \begin{align}
  \ket{\psi(t)}_\textrm{S}\propto{_\textrm{E}\langle 0|\Psi(t) \rangle_{\textrm{SE}}}.
\end{align}

 \section{dilation of Hamiltonians of two exceptional-point-based sensors}
\begin{figure}
   \centering
   \includegraphics[width=0.48\textwidth]{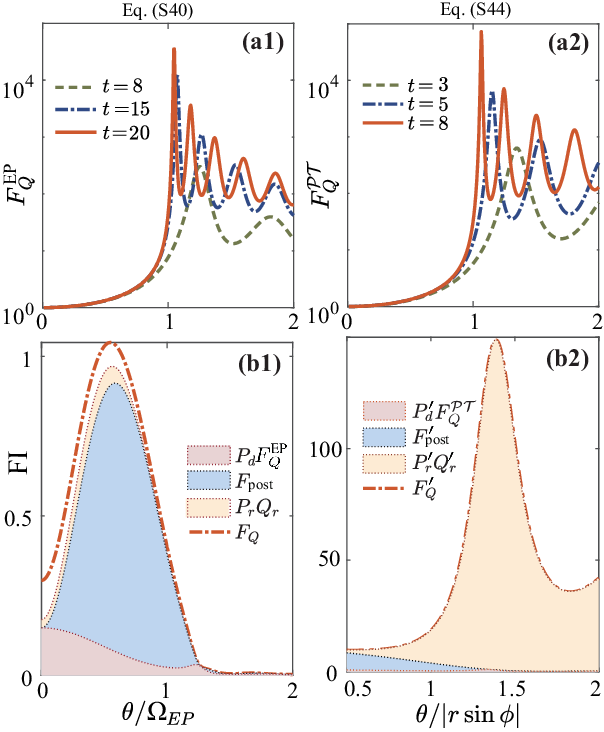}\\
   \caption{Two exceptional-point-based sensors with Hamiltonians in Eqs.~(\ref{30}) and (\ref{32}).
   (a1) The quantum Fisher information (QFI) $F_Q^{\textrm{EP}}$ for the sensor state (\ref{30}) for evolution times $t=8,15,20$, with $\Omega_{\textrm{EP}}=\omega_{\textrm{ccw}}$.
   (a2) QFI $F_Q^{\mathcal{PT}}$ for the sensor state (\ref{32}) for evolution times $t=3,5,8$, with $\phi=\pi/4$.
   (b1) and (b2) The total effective QFI $F_{\textrm{post}}$ and $F_{\textrm{post}}^{\prime}$, the effective QFI of EP-based sensors $P_dF_Q^{\textrm{EP}}$ and $P_dF_Q^{\mathcal{PT}}$ , the effective QFI for the rejected states $P_rQ_r$ and $P_r^{\prime}Q_r^{\prime}$, and the Fisher information (FI) for post-selection process $F_{\textrm{post}}$ and $F_{\textrm{post}}^{\prime}$ are shown for $t=8$ and $t=3$, respectively.
   }
   \label{fig:1}
 \end{figure}

 Here, we show more details for two examples of exceptional-point-based sensors.
 One model involves a non-Hermitian Hamiltonian that is written as
 \begin{equation}
   \begin{split}
     H^{\textrm{EP}}_\textrm{S}=\begin{pmatrix}\omega_{\textrm{cw}}&i\Omega_{\textrm{EP}}/2\\i\Omega_{\textrm{EP}}/2&\omega_{\textrm{ccw}}\end{pmatrix},\label{30}
   \end{split}
 \end{equation}
 where $\theta\equiv\omega_{\textrm{cw}}-\omega_{\textrm{ccw}}$ is the  parameter to be estimated.
 Two eigenenergies of $H^\textrm{EP}_\textrm{S}$ are
 \begin{equation}
 E_\pm\equiv[\omega_{\textrm{cw}}+\omega_{\textrm{ccw}}\pm (\theta^2-\Omega_{\textrm{EP}}^2)^{1/2}]/2,
 \end{equation}
 with $\Delta E =(\theta^2-\Omega_{\textrm{EP}}^2)^{1/2}$.
 Setting $\ket{0}_\textrm{S}$ as the initial state, the evolved state reads
 \begin{equation}
   \begin{split}
   \ket{\psi(t)}_\textrm{S}=&\left[\left(\Delta E\cos\frac{\Delta E t}{2}-i\theta\sin\frac{\Delta E t}{2}\right)\ket{0}_\textrm{S}\right.\\
   &+\left.\sin\frac{\Delta E t}{2}\ket{1}_\textrm{S}\right]/C_1,
   \end{split}
 \end{equation}
 where $C_1\!=\!(\theta^2-\cos \Delta E t)^{1/2}$ is the normalization coefficient.
 The quantum Fisher information for the state $\ket{\psi(t)}_\textrm{S}$ diverges near two exceptional points for $t\to\infty$, see Fig.~\ref{fig:1}(a1).

 Then, we use the dilation method to extend the above non-Hermitian system to a large Hermitian one.
 With Eq.~(\ref{15}), we can obtain the extended Hermitian Hamiltonian $H_{\textrm{SE}}$, and the extended system is initialized at
 \begin{equation}
 \ket{\Psi(0)}_{\textrm{SE}}=\ket{\psi(t)}_\textrm{S}\otimes\ket{0}_\textrm{E}+\hat{m}(t)\ket{\psi(t)}_\textrm{S}\otimes\ket{1}_\textrm{E}.
 \end{equation}
 From a perspective of post-selected measurements, the evolved state of the non-Hermitian system $\ket{\psi(t)}_\textrm{S}$ is the detected state $\ket{\psi_d}_\textrm{S}\equiv\ket{\psi(t)}_\textrm{S}\propto{_\textrm{E}\langle 0|\Psi(t) \rangle_{\textrm{SE}}}$, with $\ket{\Psi(t)}_{\textrm{SE}}$ and $\ket{0}_\textrm{E}$ being the joint state and the post-selection state, respectively.
 The rejected state is $\ket{\psi_r}_\textrm{S}\propto{_\textrm{E}\langle 1|\Psi(t) \rangle_{\textrm{SE}}}$.
 Then, we calculate the total quantum Fisher information $F_Q[\ket{\Psi}_{\textrm{SE}}]$ for the extended state, the effective quantum Fisher information $P_dF^{\text{EP}}_Q$ and $P_rQ_r$ for the detected and rejected states,  respectively, and the Fisher information $F_{\textrm{post}}$ for post-selection process itself, see Fig.~\ref{fig:1}(c1).

 \begin{figure}[t]
   %%\centering
   \includegraphics[width=0.48\textwidth]{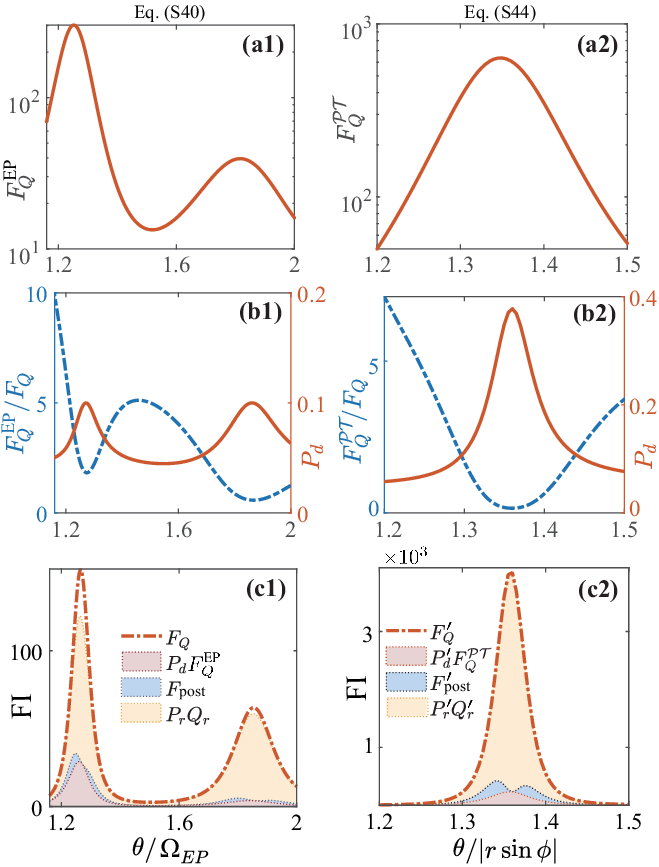}\\
   \caption{Two exceptional-points-based sensors with Hamiltonians in Eqs.~(\ref{30}) and~(\ref{32}).
   (a1) and (a2) Quantum Fisher information (QFI) for two exceptional-points-based sensors for $t=8$ and $3$, respectively.
   (b1) and (b2) Ratios of the QFI for the exceptional-points-based sensors $F_Q^{\textrm{EP}}$ and $F_Q^{\mathcal{PT}}$ to the total QFI $F_Q$ and $F_Q^{\prime}$ for the dilated Hermitian systems (blue dashed-dotted curves) are compared with the  success probabilities $P_d$ and $P_d^{\prime}$, respectively.
   (c1) and (c2) The  total QFI $F_{Q}$ and $F_{Q}^{\prime}$, the effective QFI for exceptional-points-based sensor states $P_dF_Q^{\textrm{EP}}$ and $P_dF_Q^{\mathcal{PT}}$, the effective QFI for the rejected states $P_rQ_r$ and $P_r^{\prime}Q_r^{\prime}$, and the FI for post-selection process $F_{\textrm{post}}$ and $F_{\textrm{post}}^{\prime}$ are shown for $t=8$ and $t=3$, respectively.}
   \label{fig:2}
 \end{figure}

\begin{figure}[t]
   \centering
   \includegraphics[width=0.45\textwidth]{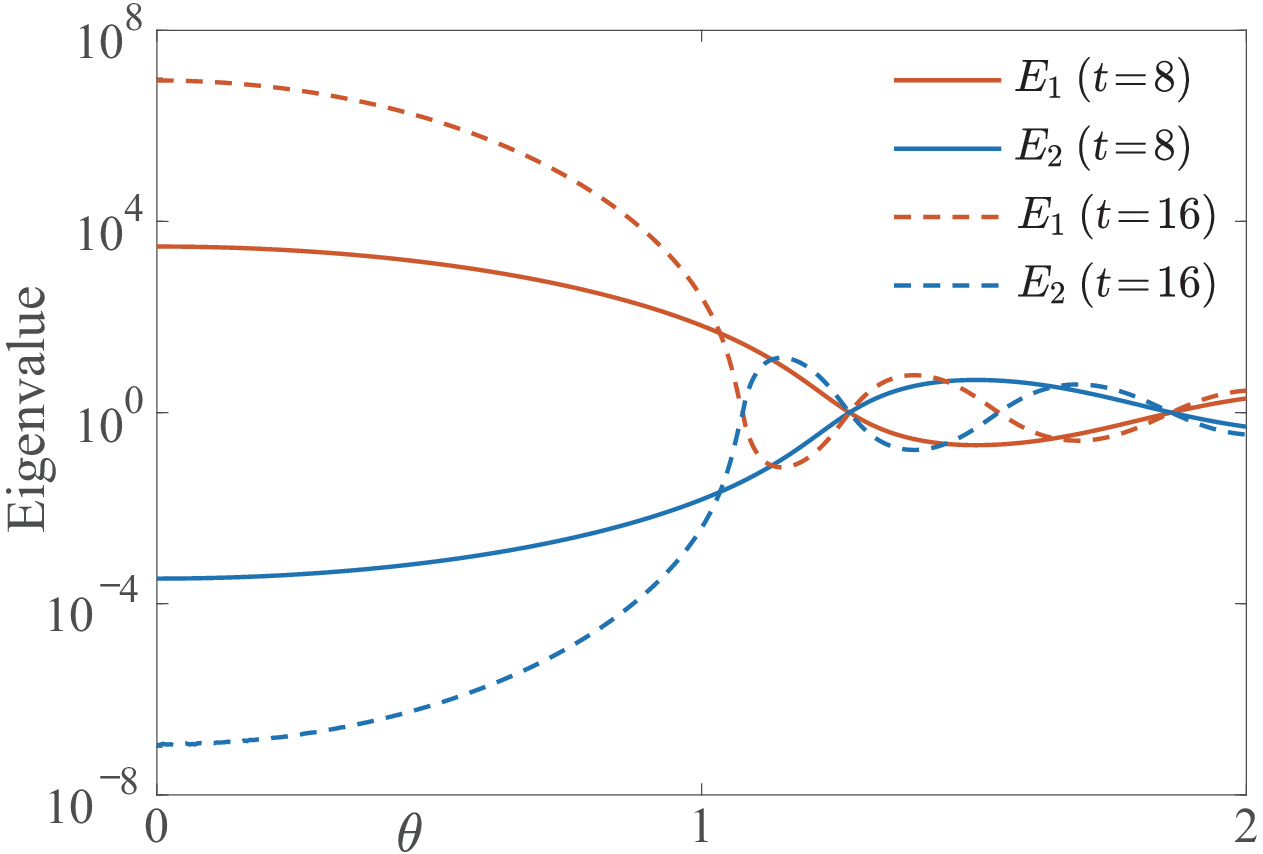}\\
   \caption{Eigenvalues $E_1$ and $E_2$ of $\hat{\eta}^{\prime}(t)$ for the model with a Hamiltonian in Eq.~(\ref{30}), for $t=8,16$ and $\hat{\eta}^{\prime}(0)={\mathbb{I}}$.
   }
   \label{fig:3}
 \end{figure}

 The other model describes a two-level $\mathcal{PT}$-symmetric system with a Hamiltonian~\cite{Yu2020}:
 \begin{equation}
     H_\textrm{S}^{\mathcal{PT}}=\begin{pmatrix}re^{i\phi}&\theta\\\theta&re^{-i\phi}\end{pmatrix},\label{32}
 \end{equation}
 where $\theta$ is the parameter to be estimated.
 Setting $\ket{0}_\textrm{S}$ as the initial state and $\phi=\pi/4$, the evoluted state reads
 \begin{equation}
   \begin{split}
     \ket{\psi(t)}_\textrm{S}\!=&\left[\left(\frac{\Delta E}{2}\cos\frac{\Delta Et}{2}\!-\!\frac{\sqrt{2}}{2}r\sin\frac{\Delta Et}{2}\right)\ket{0}_\textrm{S}\right.\\
     &\left.-\!i\sin\frac{\Delta Et}{2}\ket{1}_\textrm{S}\right]/C_2,
   \end{split}
 \end{equation}
 with $\Delta E=(\theta^2-r^2/2)^{1/2}$.
 We calculate the quantum Fisher information $F_Q^{\mathcal{PT}}[\ket{\psi}_\textrm{S}]$ as shown in Fig.~\ref{fig:1}(a2), and then consider the dilation of
 the $\mathcal{PT}$-symmetric system.
 Similarly, we calculate the total quantum Fisher information $F'_Q[\ket{\Psi}_{\textrm{SE}}]$, the effective quantum Fisher information for the detected state $P'_dF_Q^{\mathcal{PT}}$, the effective quantum Fisher information for the rejected states $P'_rQ'_r$, and the Fisher information for post-selection process itself  $F'_{\text{post}}$, as shown in Fig.~\ref{fig:1}(b2) and \ref{fig:1}(c2). In addition,
we also investigate the eigenvalues of $\hat{\eta}^{\prime}(t)$ versus $\theta$, as shown in Fig.~\ref{fig:3}.

The exceptional point ($\theta=1$) separates the phase diagram for a positive $\theta$ into two phases: the $\mathcal{PT}$-symmetry-broken phase ($0<\theta<1$) and the $\mathcal{PT}$-symmetric phase ($\theta>1$).
In the $\mathcal{PT}$-symmetry-broken phase ($0<\theta<1$), for any choice of
 $\hat{\eta}^{\prime}(0)$, one eigenvalue of $\hat{\eta}^{\prime}(t)$ will go infinity and the other approximates zero for $t\to\infty$.
 Therefore, a very large amplification factor $1/\nu^{\prime}$ is required to multiply the $\hat{\eta}^{\prime}(0)$ to ensure $[\hat{\eta}(t)-{\mathbb{I}}]$ as positive for a fixed evolution time, where $\nu^{\prime}$ is the minimum eigenvalue of $\hat{\eta}^{\prime}(t)$.
 Under this condition, the success probability $P_d$ becomes extremely small; $F_{\text{post}}$ and $P_rQ_r$ accounts for the vast majority of $F_{\textrm{tot}}$. Since most of the measurement resources are discarded, the EP-based sensor in the $\mathcal{PT}$-symmetry-broken phase is very inefficient as: $F_Q\gg P_dF_Q^{\textrm{nH}}$.
 Here, we only consider the EP-based sensor in the $\mathcal{PT}$-symmetric phase.

 Moreover, we use the time-independent dilation method to extend these two non-Hermitian systems, since this dilation method is simple, and the amplification factor $1/\nu_{\zeta}$ is much smaller than that in the
 time-dependent dilation method.
 %After that, similarly, we calculate $F_Q$, $P_dQ_d$, $P_rQ_r$, and $F_{\text{post}}$ of the two models [Fig.~\ref{fig:2}].
 %We find the EP-enhanced points of non-Hermitian sensors have an good correspondence with their extended systems, and it shows a significant improvement in the utilization rate of the measurement resources.
 The results of both examples (as shown in Figs.~\ref{fig:1} and \ref{fig:2}) verify our theory $F_Q[\ket{\Psi}_{\textrm{SE}}] \ge P_dF_Q^{\textrm{nH}}[\ket{\psi}_\textrm{S}]$, summarized as Eq.~({\color{blue}4}) in the main text.
 %However, the EP-enhanced points of non-Hermitian sensors are mediocre in the extended system.
 %Such situation is caused by the extreme parameter requirements of the extended Hamiltonian, which duo to EP.

% If we try to use an extended model to simulate an EP-system with the parameter $\theta$ covering two parts ($0<\theta<1$), whatever $\hat{\eta}^{\prime}(0)$ we set, the eigenvalues of $\hat{\eta}^{\prime}(t)$ will become divergent for $t\to\infty$ at the one of the parts (that is, one of the eigenvalues become very big $E_1\to\infty$, and the other become very small $E_2\to0$).
 %Therefore, $\hat{\eta}^{\prime}(0)$ have to multiply by a very big amplification factor $1/\nu^{\prime}$ to ensure $[\hat{\eta}(t)-{\mathbb{I}}]$ is positive for a fixed evolution time, where $\nu^{\prime}$ is the minimum eigenvalue of $\hat{\eta}^{\prime}(t)$.
 %Such extreme condition makes the success probability $P_d$ very small, and $F_{\text{post}}$ or $P_rQ_r$ will become the leading factor of $F_{\textrm{tot}}$, therefore, most of the measurement resources being ignored result in $F_Q\gg P_dF_Q^{\textrm{nH}}$.

 %Here, we only consider the $\mathcal{PT}$-unbroken side of the system to avoid the dilation at EP, witch means the side where all the eigenvalues are real for the model without $\mathcal{PT}$-symmetric.

 \begin{figure}
  \centering
  \includegraphics[width=0.46\textwidth]{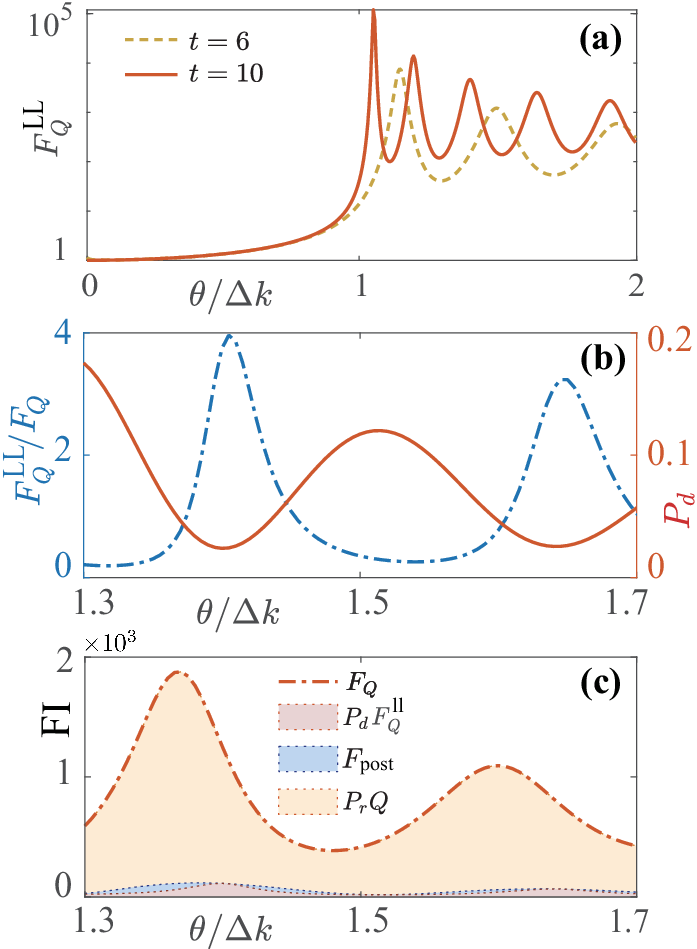}\\
  \caption{Loss-loss $\mathcal{PT}$-symmetry sensor with a Hamiltonian
  (\ref{34}) and $v=g=1$.
    (a) Quantum Fisher information (QFI) for the loss-loss sensor state for  $t=6$ and $10$.
    (b) Ratio of the QFI for the loss-loss sensor $F_Q^{\textrm{LL}}$ to the total QFI $F_Q$ for the Naimark dilated system (blue dashed-dotted curve) is compared with the success probabilities $P_d$ (red solid curve).
    (c) Total QFI, $F_Q$, the effective QFI for the sensor state $P_dF_Q^{\textrm{LL}}$, the effective QFI for the rejected state, $P_rQ_r$, and the Fisher information (FI) for post-selection process, $F_{\textrm{post}}$.}
  \label{fig:4}
\end{figure}

 \section{model of loss-loss $\mathcal{PT}$-symmetry sensor}
   Finally, we consider a loss-loss (LL) $\mathcal{PT}$-symmetry sensor with a Hamiltonian as experimentally implemented in Ref.~\cite{ruan2025}:
 \begin{equation}
   H_\text{LL}=\begin{pmatrix}v-ik_\textrm{H}&ig\theta\\-ig\theta&v-ik_\textrm{V}\end{pmatrix},\label{34}
 \end{equation}
 where $\theta$ is the unknown parameter to be estimated.
 Choosing $\ket{\psi(0)}$ as the initial state, the time-evolved state is
 \begin{equation}
   \ket{\psi_\text{LL}(t)}=e^{-iH_\text{LL}t}\ket{\psi(0)}.
 \end{equation}
The Hamiltonian  $H_\text{LL}$ can be divided into two terms
 \begin{equation}
   \begin{split}
     H_\text{LL}&=\begin{pmatrix}v-i\Delta k&ig\theta\\-ig\theta&v+i\Delta k\end{pmatrix} -\begin{pmatrix}ik&0\\0&ik\end{pmatrix}\\
     &=H_{\textrm{GL}}-ik{\mathbb{I}},
   \end{split}
 \end{equation}
 where $k=(k_\textrm{V}-k_\textrm{H})/2$, $\Delta k=(k_\textrm{V}-k_\textrm{H})/2$, and $H_{\textrm{GL}}$ denotes the Hamiltonian of a gain-loss (GL) non-Hermitian system.
 After a gauge transformation, the evolution can be represented as
 \begin{equation}
   \ket{\psi_{\textrm{LL}}(t)}_\textrm{S}=e^{-iH_{\textrm{LL}}t}\ket{\psi(0)}_\textrm{S},
 \end{equation}
 with $\ket{\psi_{\textrm{LL}}(t)}_\textrm{S}=\exp(-kt)\ket{\psi_{\textrm{GL}}(t)}_\textrm{S}$.
 That is, the dynamics of a loss-loss system is mathematically equivalent to that of a gain-loss system under the impact of a global exponential decay.
 Nevertheless, the loss-loss system can avoid the gain-induced noise.

 Since the global exponential decay is independent of $\theta$, we only focus on the $H_{\textrm{GL}}$.
 The eigenenergies of $H_{\textrm{GL}}$ are $E_{\pm}\equiv v\pm (\theta^2-\Delta k^2)^{1/2}$, with the energy difference as $\Delta E=2(\theta^2-\Delta k^2)^{1/2}$, which vanishes as two exceptional points ($\theta=\pm \Delta k$).
 By setting $\ket{0}_\textrm{S}$ as the initial state, the time evolved state under $H_{\textrm{GL}}$ is
 \begin{equation}
  \ket{\psi_{\textrm{GL}}(t)}_\textrm{S}=[(a\cos at - \Delta k \sin at)\ket{0}_\textrm{S}-\theta \sin at \ket{1}_\textrm{S}]/C_{\textrm{GL}},
 \end{equation}
with $C_{\textrm{GL}}=\theta^2- \Delta k^2 \cos 2at -a \Delta k  \sin 2at$, and $a=\Delta E/2$.

 The quantum Fisher information for the evolved state $F_Q^{\textrm{LL}}[\ket{\psi_{\textrm{LL}}(t)}_\textrm{S}]$ can be arbitrary large for $t\to\infty$ near the exceptional points ($\theta/\Delta k=\pm1$), which is shown in Fig.~\ref{fig:4}(a). 
 Similarly, by extending the non-Hermitian system with the Naimark dilation method, we consider the ratio of the quantum Fisher information for the non-Hermitian sensors $F_Q^{\textrm{LL}}$ to the one for  the extended Hermitian system $F_Q$, which are compared with the post-selection success probability $P_d$, see Fig.~\ref{fig:4}(b).
 In addition, we consider the effective quantum Fisher information of the non-Hermitian sensor, which is smaller than the one of its Hermitian counterpart as $P_dF_Q^{\textrm{LL}}<F_Q$ [Fig.~\ref{fig:4}(c)].

 From Fig.~\ref{fig:4}(b) and \ref{fig:4}(c), the effective quantum Fisher information of the non-Hermitian sensor $P_dF_Q^{\textrm{LL}}$ achieves the maximum when the post-selected probability $P_d$ achieves its local minimum.
 It means more information of $\theta$ can be obtained with less trials of measurements using the non-Hermitian sensor from its Hermitian counterpart, i.e., "when less is more".
 However, the effective quantum Fisher information  $P_dF_Q^{\textrm{LL}}$ of the non-Hermitian sensor is still much smaller than the one for its Hermitian counterpart cause the loss of information during the post-selection process.
 Our investigation employs quantum measurements with a weak probe, which inherently yields a low success probability. In contrast, classical measurements using a strong probe may achieve a higher success probability, presenting a problem for future research.

%	\emph{Conclusions.---}%

\bibliography{Supplemental}